# Strategically-Motivated Advanced Persistent Threat: Definition, Process, Tactics and a Disinformation Model of Counterattack

**Atif Ahmad**
School of Computing and Information Systems
The University of Melbourne
Parkville, Australia
atif@unimelb.edu.au

**Jeb Webb**
Oceania Cyber Security Centre
Melbourne, Australia
jeb.webb@ocsc.com.au

**Kevin C. Desouza**
QUT Business School
Queensland University of Technology
Queensland, Australia
kevin.desouza@qut.edu.au

**James Boorman**
Oceania Cyber Security Centre
Melbourne, Australia
james.boorman@ocsc.com.au

## Abstract

Advanced persistent threat (APT) is widely acknowledged to be the most sophisticated and potent class of security threat. APT refers to knowledgeable human attackers that are organized, highly sophisticated and motivated to achieve their objectives against a targeted organization(s) over a prolonged period. Strategically-motivated APTs or S-APTs are distinct in that they draw their objectives from the broader strategic agenda of third parties such as criminal syndicates, nation-states, and rival corporations. In this paper we review the use of the term "advanced persistent threat," and present a formal definition. We then draw on military science, the science of organized conflict, for a theoretical basis to develop a rigorous and holistic model of the stages of an APT operation which we subsequently use to explain how S-APTs execute their strategically motivated operations using tactics, techniques and procedures. Finally, we present a general disinformation model, derived from situation awareness theory, and explain how disinformation can be used to attack the situation awareness and decision making of not only S-APT operators, but also the entities that back them.

## 1.0 Introduction

- On 3 October 2018, FireEye published an article on what is thought to be a state-sponsored advanced persistent threat (APT) team dubbed "APT38" (Fraser et al 2018). According to FireEye, APT38 have been "active since at least 2014" and involved in theft estimated at more than "a hundred million dollars" from banks across 11 countries so far (Fraser et al 2018). The team are thought to be well resourced and organised, using custom tools to gain long-term access ("on average.. 155 days.. up to 2 years") to targeted financial organisations and avoid detection while gathering intelligence ("network layout, required permissions, and system technologies") to enable a successful attack, followed by deleting logs or deploying "disk-wiping malware" to compromised systems to cover their tracks (Fraser et al 2018).

- As of 5 September 2018, the still-active advanced persistent threat (APT) team dubbed "Silence" has used known tools in combination with its own custom-made toolkits to steal the equivalent of over USD $800,000 from banks, online stores, insurance companies, and news agencies in Azerbaijan, Belarus, Kazakhstan, Poland, Russia, and Ukraine (Vijayan 2018). The team, believed to consist of only an operator who has "in-depth knowledge about tools for conducting [penetration] tests on banking systems, navigating inside a bank's network, and gaining access to protected systems" and a developer who "seems to be an adept reverse-engineer who is responsible for developing the tools and exploits," averages one attack every three months. Group-IB, who has been tracking Silence, has assessed that the team are likely information security practitioners who draw on current intelligence about systems, security technologies, threats and vulnerabilities (Vijayan 2018).

- On 3 August 2018, the Singapore government announced that its health system had been compromised in what it believed to be a state-sponsored attack which involved the exfiltration of politician's medical records (Ng 2018). The team responsible for the APT attack used custom malware to circumvent the security technologies in place and had been able to maintain undetected access for some time. Threat intelligence analyst, Joanne Wong, of LogRhythm suggested that the "health records of Singapore's leadership could be used to cause instability within the country by casting doubts on the health status of our leaders" (Choo 2018).

Organizations operate within a highly complex, ever evolving, and sophisticated security threat landscape that exposes their digital assets (data, algorithms, models, analytics, systems, and infrastructure) to a myriad of security risks (Shedden et al., 2016; Shedden et al., 2011). The information security management function serves to protect the organization from security threats through the application of formal, informal and technological controls (Ahmad and Maynard, 2014). Probably the most potent security threat to organizations is the so-called APT or "advanced persistent threat" (Kim et al. 2014; Maisey 2014; Moon et al. 2014; Nelson 2014).

APTs are perpetrated by technological experts who are well-trained, well-funded, organized, and capable of utilizing a range of technologies to achieve their objectives over a prolonged period (Hutchins et al. 2011; Maisey 2014; Mansfield-Devine 2014). Over the last decade there have been numerous media reports and much literature devoted to the APT. In the period 2016 to 2018, APTs have been implicated in financial crime (Hern 2016; Murray and Dyson 2016; Constantin 2017), political and industrial espionage (Ashok 2017; Choudhury 2017), and the attempted influencing/sabotaging of democratic election processes (Thielman 2016; Graham 2017).

Given the rising significance of APTs, it is unfortunate that the information security literature remains fragmented, which limits our ability to build cumulative knowledge about the nature of these threats, their governing dynamics from an operational perspective, and design innovative solutions to combat them. Towards this end, this paper adopts an 'organized conflict' perspective on APTs, providing a more holistic view of the APT phenomenon by putting APT tactics in the context of operations—an operation being "a sequence of tactical actions with a common purpose or unifying theme" (JP 1-0 2013, I-9). It is a novel socio-organizational approach to understanding APTs that makes contributions to management strategy not previously possible when adopting a purely technology-centric (i.e. technical measure/technical countermeasure) perspective.

The rest of the paper is structured as follows. First, we offer a brief background on organized conflict from the perspective of military doctrine, to include the role of human situation awareness and decision-making in tactical and strategic maneuver. Next, we describe our research methodology. This is followed by an overview of the origins and uses of the term "advanced persistent threat," after which we present our own formal definition of the term. We subsequently define a special variant of APTs known as strategically-motivated APTs or S-APTs. Using a simple typology, we distinguish S-APTs from other APTs in that S-APTs draw their objectives from the broader strategic agenda of third parties. We then present an operational framework for interpreting advanced persistent threats: the APT Operation Line (APTOL) model, which we subsequently use to structure a rich discussion of strategically motivated APT tactics, techniques and procedures (TTPs) reported in the literature. This is followed by an examination of the role of human situation awareness implicit in these operations, and how it can be exploited as a vector for counterattack. A disinformation model based on Endsley's (1995) theory of situation awareness is presented to illustrate how to counter S-APT operations and undermine the broader strategic interests of such operators (the teams executing the attack) and their backing entities (the entities setting the strategic objectives of the attack). We conclude the paper with a discussion of the implications of our framework for researchers and practitioners.

## 2.0 Background

The domains of military science and situational awareness (Endsley 1995) are useful toward understanding APTs for several reasons. First, APTs constitute planned and rigorously organized attacks on the digital assets of a targeted organization. Second, to combat APTs, organizations need to increase, and leverage, their situational awareness on APTs and the evolving threat landscape. Third, the success of either party will come down to the art and science through which information-driven strategies are employed.

### *Organized conflict: a military science perspective*

Military doctrine presents cumulative, high-level findings from a long history of integrated research, analysis, development, testing, practical application, and lessons learned (Piehler 2013, 884; Klein 1989; Van Creveld 1985). As military science is the science of organized conflict, it provides a useful lens for interpreting, defining and characterizing APTs, which involve organized activities carried out in support of objectives that may extend well beyond the attack's immediate consequences for data and systems, or even the predictable negative consequences for people and organizations. These kinds of operations can be strategic instruments for change, through which wealth and power are gained in stages at the expense of other people, organizations, societies and nations. Ultimately, APT operations represent organized conflict situations in which organized entities apply resources in deliberate ways to attack or defend assets. It is warfare, but typically not recognized as such under the provisions of national and international law (i.e. given that specific legal criteria must be met for belligerent acts to formally constitute "acts of war").

APTs can be viewed from three perspectives: 1) a desired big-picture end state, 2) a general plan for getting there (which shapes the situation as it unfolds), and 3) the specific steps involved in carrying out said plan. The inducement of cumulative effects to achieve strategic objectives is well-covered in military doctrine. In military parlance, the big-picture end state, the plan for getting there, and the steps involved are referred to as the strategic, operational and tactical levels of war, respectively (JP 1 I-7—8, 2013). As defined in JP 1-0, Doctrine for the Armed Forces of the United States, "An operation is a sequence of tactical actions with a common purpose or unifying theme" (JP 1-0 2013,

I-9). The "common purpose" behind this "sequence of tactical actions" is the motivation for the operation—the "end states" or "strategic objectives" that the operation ultimately aims to achieve (I-8—9). APTs are operations—what the US military currently refers to as "offensive cyberspace operations (OCO)" (JP 3-12 2013, II-2).

The immediate goals achieved through each of these operations build toward the achievement of broader strategic objectives for APT operators or any powers who might back them. For example, while successfully stealing the formula for an innovative product might achieve the goal of the operation targeting one company; the theft of many different types of intellectual property from many different companies within the same industry could enable the achievement of a backing power's strategic objective of dominating that market. Understanding that the APT is an operation is the key to its defeat. Operations are goal-oriented undertakings, bounded in terms of the limited number of tactical movements available to the operator for achieving one or more specific objectives. The operational picture links tactics to the achievement of operational goals that build toward strategic objectives. Reverse-engineering the attacker's concept of operations can enable the defender to anticipate tactical movement, apply resources effectively, and disrupt the operation. Operations can be disrupted to deny the adversary's strategic objectives or can even be manipulated (e.g. through intentional misinformation/disinformation) to undermine the adversary's broader strategic interests.

### *Situation Awareness*

Situation awareness—awareness of what is going on within some situation of interest—is necessary for informed decision-making and appropriate action (Smith and Hancock 1995; Endsley 1995; Bedny and Meister 1999). In this paper we employ Endsley's theoretical model of Situation Awareness because it outlines a linear process which lends itself easily to specification of a "counter process" intended to deny situation awareness (Endsley 1995).

Endsley (1995) models the SA development process in three progressive stages that range from relatively low awareness to relatively high awareness regarding a situation. Generically speaking, "situations" consist of objects or phenomena that exist within the external "environment" (as opposed to matters originating within the internal environment of the mind). SA concerns a cognizing agent's awareness of the circumstances under which goal-oriented decision and action occur. First, detectable elements representing some object or phenomenon external to the cognizing agent are detected via one or more sense faculties and cognitively registered, thereby establishing a state of basic "perception" (Level 1 SA). Through the comparison of sensed patterns with structures in memory, recognition regarding the sensed object or phenomenon is achieved, along with an imputation of meaning that marks a state of "comprehension" (Level 2 SA). If the cognizing agent can comprehend the relevant elements of a situation, the relationships between them, and the context within which they exist, to such an extent that implications can be deduced or predicted reliably, then "projection" (Level 3 SA) has been achieved.

At the tactical level, situation awareness is required to ensure that actions result in desirable effects at instances (JP 2-0 2013, I-25). At the operational level, situation awareness is required to ensure that all tactical actions are coordinated and conducive to the achievement of operational goals, i.e. a desirable end state that is worked toward across all tactical instances occurring within the defined operational context (JP 2-0 2013, I-24). At the strategic level, situation awareness is required to ensure that the aggregate end states of operations are conducive to the attainment of some desirable relative status within the context of some greater competitive environment (JP 2-0 2013, I-23—24).

In the context of an APT operation, the APT operator must maintain situation awareness within the operational environment to ensure that tactical actions (i.e. intelligence collection, movements, maneuvers and fires) are appropriate and effective, and that they are conducive to force protection/operational security (FM 3-90-2 2013, I-3; FM 3-38 2014, 3-11). Furthermore, situation awareness is required at the operational level to link tactics to operational objectives and operational objectives to strategic objectives (JP 3-0 2011, I-12—14). Such coordination necessarily requires an understanding of the commander's intent (JP 1 2013, V-15,) and may require the feeding forward of intelligence for command-level/strategic decisions (JP 1 2013 V-19; FM 3-90-2 2013, I-3). While the degree of autonomy afforded to APT operators may vary across circumstances, some degree of integrated situation awareness across tactics, operations, and strategy is required wherever a chain of command exists (JP 2-0 2013, 1-4; FM 3-90-2 2013 1-1).

As awareness of a situation improves, the capacity for "appropriate" decision and action (i.e. decision and action that will result in a desired outcome, given the reality of a situation) increases. As awareness decreases, this capacity naturally decreases as well. SA is a virtual condition. It is essentially an internal/subjective representation of an external/objective reality, based on the available evidence. Changes to the available evidence therefore result in changes to SA. Where there is an understanding of the cognizing agent's information structures in memory, evidence can be selected (or designed) to induce changes to that agent's mental model of a situation. The act of intentionally making disinformation available to a cognizing agent, so that the SA development process results in a false but seemingly credible representation of reality in the mind of that agent, is the act of deception explained in terms of SA theory.

## 3.0 Research Methodology

We conducted an exhaustive search spanning several literatures to assemble a comprehensive description of APTs. First, an open search was conducted using popular literature databases (via the EBSCO interface) for the term "advanced persistent threat" to establish an early chronology of its usage in public discourse. Next, an "all text" search (limited to peer-reviewed publications in the English language) was conducted for the phrase "advanced persistent threat" using the same literature databases. After removal of duplicate entries, this search yielded 110 results.

The results were then manually checked for relevance, and the pool was narrowed to 38 sources that went beyond mention of the term APT to provide some discussion of attributes characteristic of the threat type (i.e. the other sources mentioned APTs without going into substantial detail about them). Of these 38 sources, 14 characterized APTs as structured processes consisting of definite steps. We then isolated the three journals that yielded the most returns within the initial 110 results—Network Security (23); Computer Fraud & Security (15); and Computers & Security (6)—and performed another search within these titles for the acronym "APT". The rationale for doing this was that, given the familiarity of the subject matter to their readership, some authors might use the acronym without elaboration in the body of the article. This search yielded four additional relevant articles (two offering process descriptions), bringing the total to 42 substantial treatments (with 16 process descriptions).

The articles were analyzed using a pattern-matching illustrative method (Neuman 2014), which involved iteratively comparing APT tactics described in the literature against JP 3-12's (2013) definitions of the six complementary functions which must be integrated into any operation for that operation to be successful. The output of this process was a process model in the form of the APT Operation Line framework (offered in section 6 of this paper). The stages or "tactical milestones" of this model were then used to organize the literature review portion of this paper, which focused on collecting information about tactics used by APT operators. (While our own review of the literature was exploratory in nature—to identify a knowledge gap and an opportunity to contribute to theory and practice—our treatment of the literature here has been structured this way simply to demonstrate the explanatory utility of the model). Noting that the treatments of APTs in the literature generally focus on their technological dimension, we have focused instead on their human dimension. Viewing the tactics applied in APTs as goal-oriented actions taken by operators within operational environments, Endsley's (1995) situation awareness theory was repurposed (put into opposite terms) to provide a theoretical basis for disinformation measures intended to adversely impact situation awareness, decision-making, and action by APTs and their backing entities (see section 8 of this paper). Non-peer reviewed sources are included in some portions of the literature review to provide more specific details on tools, tactics, and procedures.

## 4.0 APTs: Origin and Evolution of the Concept

Within the public domain, the term "advanced persistent threat" first appears in a US patent application filed in 2007 and published in 2008 (Schmidt et al. 2008, 3), which describes the threat:

> *It is characterized by greater sophistication and skill, rapid collaboration, and increasingly structured relationships to overwhelm complex network security mechanisms—oftentimes from the inside. Their motivation is becoming increasingly profit-focused, and their modus operandi includes persistence and stealth. It includes possible state-sponsored actors whose effects contribute to long-term influence and exploitation campaigns, as well as devastating effects to facilitate military action. Their signatures include the use of zero-day exploits, distributed agent networks, advanced social engineering techniques such as spear phishing, and long-term data mining and exfiltration. Their flexibility and robust kitbag of tools and techniques makes the advanced threats particularly difficult to successfully defeat with today's technology-heavy network security focus.*

While the patent includes "possible state-sponsored actors" in its description of the APT, even at this early stage, the term was already being used semi-generically to describe a threat scenario as opposed to an attacker. In mainstream news media, on the other hand, the term APT was originally employed for specific reference to state-sponsored actors (Walsh 2008; Grow et al. 2008; Bejtlich 2010). China being the state most often associated with it (Bradbury 2010; Messmer 2011).

The first instance of APT linked to state-sponsored cyberattack appears in an interview with Northrop Grumman CISO Timothy McKnight, where McKnight explains that it is a term that the US Department of Defense uses to refer specifically to "the nation-state threat," or "country-sponsored attacks" that constitute "well-resourced, highly targeted attacks at corporations and governments [by groups] that are looking primarily to steal intellectual property and gain competitive advantage" (Walsh, 2008, 13). Two months later, Businessweek also refers to APTs as "a new form of attack" that is state-sponsored, and which involves "using sophisticated technology" (Grow et al. 2008, 35). Bejtlich (2010, 21) claims that the United States Air Force "coined the phrase advanced persistent threat in 2006," and that the term is used to refer to "specific threat actors," albeit in generic terms, for a mixed audience that includes people who do not have national security clearances. "APT does not refer to vaguely unknown and shadowy Internet forces. The term is most frequently applied to distinct groups operating from the Asia-Pacific region" (Bejtlich 2010, 21).

By the end of 2010, "APT" was not being applied strictly to state-sponsored operations within the information security literature. APT was used to refer to long-term hacking campaigns by well-resourced adversaries against specific targets using sophisticated tactics, techniques and procedures or TTPs (e.g. see Chen et al., 2014; Ussath et al., 2016; NIST SP 800-39). Although the fact that APTs are conducted for powerful entities with something to gain is frequently acknowledged in this literature, however, the implications of this fact have not been thoroughly explored. Much of the discussion has centered on a technology-centric perspective of tactics, with varying degrees of effort applied toward relating these tactics to one another in the context of a definite process or lifecycle.

From the articles we collected, 14 presented APT using a process approach. Brewer (2014), Moon et al. (2014), Virvilis et al (2014), and Wangen (2015) all offer similar process descriptions that include learning about the target organization and network, attacking/penetrating the network, enabling persistent access, and stealthily collecting information on an ongoing basis for as long as possible or necessary. The earliest of these reproduce models that were originally published in non-peer-reviewed publications[1], i.e. the model described by Cutler (2010) appears in Tankard (2011); and the Security for Business Innovation Council's (2011) appears in Dorey and Leite (2011). Hutchins et al. (2011)—which, according to Google Scholar is now the single most-cited source on APTs—does not appear to be referenced in peer-reviewed literature until Maisey (2014).

Existing process descriptions of APTs (Tankard 2011; Dorey and Leite 2011; Smith 2013; Sood and Enbody 2013; Brewer 2014; Kim et. al 2014; Moon et al. 2014; Steer 2014; Virvilis et al. 2014; Maisey 2014; Wangen 2015; and Chang et al. 2016), are presented 'as-is' without any theoretical basis, comprehensive explanation or description of their development from first principles. Moreover, process descriptions (see Brewer (2014), Moon et al. (2014), Virvilis et al (2014), and Wangen (2015) only adequately describe APT missions aimed at obtaining information. They are not as applicable to APT missions oriented toward sabotage (of systems, processes, or operations), such as missions that culminated in the deployment of Stuxnet, Shamoon 2, or Crash Override exploit kits. Stuxnet, which was used to target Iran's uranium enrichment program in an automated attack attributed to US and Israeli developers, was designed to induce control system routines that caused centrifuges to rapidly deteriorate (Ivezic 2018). Shamoon 2, which was used to target Saudi Arabian companies and government agencies in attacks attributed to Iranian developers, was designed to destroy computers by erasing data from their hard drives (Fixler 2018). Crash Override, which was used to target the Ukrainian power grid in an attack attributed to Russian developers, was designed to shut down various forms of equipment via control system commands (Greenberg 2017).

The security industry appears to have a similar view of the high-level process of APTs to information security researchers, where organizations are targeted with sophisticated tools to attack/penetrate the network, enable persistent access, and stealthily collect information on an ongoing basis for as long as possible or necessary to achieve a goal (Symantec 2011; Check Point 2019; CISCO 2019; FireEye 2019; Kaspersky Lab 2019; NIST SP 800-39). Both NIST and Symantec detail the additional step of going beyond data exfiltration, including the potential goal of disruption or sabotage (Symantec 2011; NIST SP 800-39). A statement from ESET in their analysis of the Lojax UEFI root kit, believed to be developed by the Sednit group (also known as APT28), suggests that APTs are essentially just tool sets that can be used by different actors to achieve a variety of goals (ESET 2018). ESET note: "What we call "the Sednit group" is merely a set of software and the related network infrastructure, which we can hardly correlate authoritatively with any specific organization." (ESET 2018).

None of the peer-reviewed APT process descriptions capture the human dimensions of awareness and decision-making involved. However, an industry white paper from Symantec references an opinion article by Brenner (Brenner 2011) and alludes to the strategic nature of some APTs when they suggest that nation-states are actively targeting organizations with "valuable technology or intellectual property" as "national security and economic security have converged" (Symantec 2011). In support of this view, CISCO's discussion of APTs suggest that APTs may be launched by nations with "…significant political motivations, such as military intelligence" or by others groups to obtain "…significant competitive advantages or lucrative payouts." (CISCO 2019).

## 5.0 A formal APT definition

The term 'Advanced Persistent Threat' was originally intended to describe malicious, organized, and highly sophisticated cyber campaigns where the strategic objectives came from an external backing entity. However, our review of the literature shows that information security researchers and industry professionals tend to use the term APT more broadly to address malicious, organized and highly sophisticated cyber campaigns in general. Using a simple typology we identify the characteristics that the literature does agree upon – specifically that APTs are a type of cyber threat that are malicious, organized, highly sophisticated in their use of TTPs and target IT networks in specific organizations (to obtain information or sabotage its operations) for long-term access:

---

[1] However, because of their central roles in the citing peer-reviewed works, some of these original sources are also cited elsewhere in this paper.

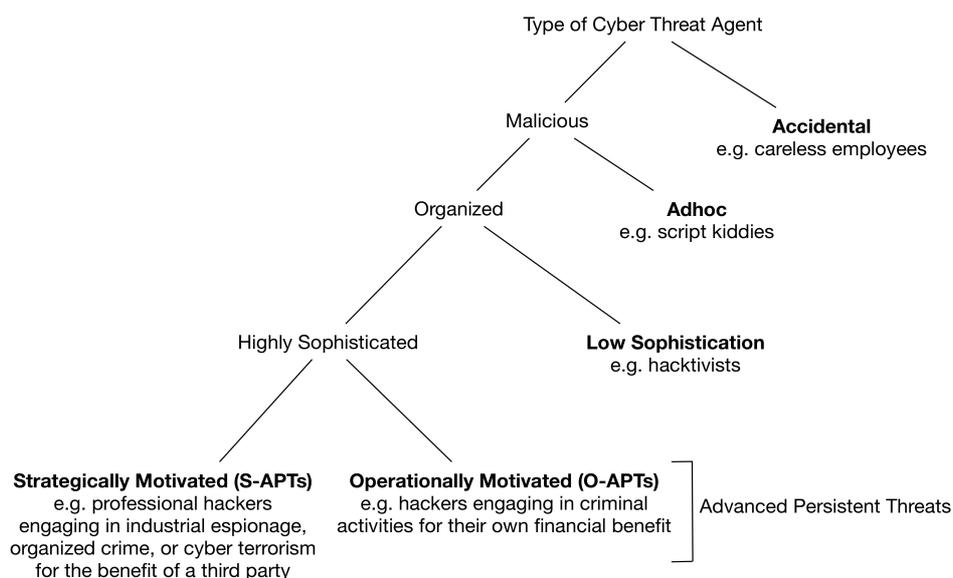

**Figure 1: A Typology of Cyber Threat Agents**

Therefore, towards building a consensus on the definition of an APT, we present the following in formal language that has utility to both researchers and practitioners:

> *An entity that engages in a malicious, organized, and highly sophisticated long-term or reiterated network intrusion and exploitation operation to obtain information from a target organization, sabotage its operations, or both.*

As the definition above states, APT operations serve one or both of two core functions: obtaining information (through data theft or intelligence collection) or sabotage[2], i.e. "capabilities that can be employed to deceive, degrade, disrupt, deny, destroy, or manipulate across the continuum" (FM 3-38 2014, 1-9)[3].

## *Obtaining Information*

Intelligence collection is always necessary for an operation and is sometimes the sole objective of an operation (JP 2-0 2013, I-11). Conducting an operation to obtain certain types of information must also involve the collection of other types of information necessary for planning and executing the operation itself. This kind of intelligence support to operations is generally achieved through reconnaissance activities.

Reconnaissance is "the systematic observation of places, persons, or things" (JP 2-0 2013, I-11), which is to say that the observation is continuous over an extended period. Reconnaissance involves a) evaluating enemy presence and identifying the enemy's offensive capabilities; b) defeating offensive capabilities where possible/feasible and authorized; c) mapping objects in terrain and determining avenues for movement (to include identification of obstructions to movement and no-go zones); and d) reporting all relevant information back to the decision-maker directing the operation (FM 3-90-2 2013, I-9). Interlocking or overlapping reconnaissance missions can achieve surveillance.

## *Sabotage*

In contrast to forms of cyber-attack that have immediate apparent effects (such as denial of service), sabotage operations are often insidious (Jasper 2015). Sabotage can be orchestrated to induce indirect, cumulative, cascading, or otherwise unexpected effects for information systems (to include human decision-making) and the people and processes supported by them (JP 3-60 2007, I-10—11; FM 3-38 2014, 3-3).

A covert offensive operation focused on the inducement of harmful effects[4] for the target that in turn support the attacking side's objectives, sabotage can be modeled as a six-phase cycle: 1) establishing the desired goal; 2) target selection informed by target role/systemic context (to critically assessing the appropriateness of the target, given the desired goal); 3) assessing the capacity for available fires/inducement techniques to affect the target in the desired way; 4) command-level decision and direction; 5) planning and execution of specific target engagements (i.e. applying the F2T2EA "kill chain:" find, fix, track, target, engage, assess); and 6) assessing whether the desired goal has been achieved (FM 3-60 2007, II-3—19).

---

[2] "An act or acts with intent to injure, interfere with, or obstruct…" (JP 1-02 2010, 209); used here to refer to any inducement of effects intended to undermine the target's ability to achieve goals.

[3] The strategic objectives of a CNE operation may include both information obtainment and sabotage.

[4] "Many of the ways and means associated with targeting and employing fires result in tactical-level effects relative to the selected targets. However, the cumulative results of these target engagements can contribute to…desired operational-level and theater-strategic effects" (JP 3-60 2007, I-8—9).

*Theoretical Underpinnings for Operations*

JP 3-0, Joint Operations, (2011) explains that six functions are common to all operations: command and control (C2), intelligence, fires, movement and maneuver, protection, and sustainment (III-2—III-39). Military doctrine holds that these "…functions reinforce and complement one another, and integration across the functions is essential to mission accomplishment" (JP 3-0 2011, xiv). Command and control is the direction of activities in accordance with the will of the commander, i.e. the strategic decision-maker who has ultimate decision-making authority (III-2). Intelligence is "integrated, evaluated, analyzed, and interpreted information" about the enemy or the operational environment (III-20). "Fires" refers to the use of weapons or systems to effect changes in the state of a target (III-22). Movement and maneuver refers to the combination of movement and fires to achieve an advantageous position (III-28). Protection and sustainment refer to safeguarding one's own force and assuring continuity of operations, respectively (III-29, III-35).

JP 3-12(2013) explains how these six functions apply to cyberspace operations (CO):

- Command and Control - A commander's control over forces tasked with carrying out CO (II-6).
- Intelligence - Intelligence that is pertinent to the conduct of CO as well as any intelligence that can be conducted via CO (II-8).
- Fires - The use of "cyberspace capabilities" to "manipulate adversary cyberspace targets through (military deception) redirection, systems conditioning, etc." (II-10).
- Movement and Maneuver – The movement of data through physical and logical infrastructure and the navigation of network links and nodes, often in conjunction with the application of fires (II-10—11).
- Protection – Protection of physical and logical infrastructure, and the application of defensive capabilities and operational security (sometimes equivalent to information security) measures (II-12).
- Sustainment – Maintaining capabilities through equipment acquisition, training, capability upgrades, and planning for operational continuity (II-11).

The above six operational functions, and the preceding fundamentals and tasks associated with reconnaissance and the phases and steps associated with attack/sabotage operations, can be combined with findings reported in the literature to produce an APT operation line model, based on the line of operation concept put forth in US military doctrine (JP 5-0 2011, 3-27), which links the tactical milestones composing an operation to the strategic objectives of that operation.

As APT operations are, like any operation, human-driven processes carried out to fulfill human objectives, situation awareness is required at all decision points: tactical, operational and strategic. The situation awareness dimension should be included in a description of the APT's line of operation because, as we will show later in this paper, vulnerabilities inherent to this dimension can be exploited by defenders to deceive adversaries, degrade their capabilities over the short and long terms, disrupt their operations and strategic agendas, deny them strategic advantage and, potentially, even destroy their information assets or manipulate their planning processes in disadvantageous ways that result in strategic failure.

# 6.0 An Operational Framework for Interpreting the Objectives of Strategically-motivated Advanced Persistent Threat

Given our research interest lies in strategic decision-making in organizations, we focus the remaining part of the paper on APTs where the attacking entity or operator draws their strategic objectives from an external party (see vignette 1 on APT38 and vignette 3 on the compromise to Singapore's health system in the introduction to this paper). We call these 'Strategically-motivated Advanced Persistent Threat' or S-APTs. S-APTs are similar to Operationally-motivated APTs or O-APTs (see our typology in figure 1 to distinguish between the two) in that they use the same types of TTPs, however they develop, use, combine, and integrate TTPs to serve the strategic purposes of a third party (see vignette 2 on the "Silence" APT for an example of an apparent O-APT).

We define S-APTs as follows:

> *An entity that engages in a malicious, organized, and highly sophisticated long-term or reiterated network intrusion and exploitation operation to obtain information from a target organization, sabotage its operations, or both,* <u>in support of that or another entity's broader strategic agenda to gain or maintain wealth or power</u>.

Acknowledging the strategic motivation of S-APTs is critical for organizations that are building defensive systems against highly sophisticated and persistent actors. At a tactical level understanding the attacker's strategic objectives can enable the defender to anticipate tactical movement, apply resources effectively, and disrupt the operation. At a strategic level it is critical to target the strategic situation awareness and decision-making of the backing entity (e.g. convincing the backing entity that the company has a different business model than it actually does, that its R&D program has a different trajectory than it actually does; or that it uses or depends on systems differently than it actually does). Since the backing entity cannot be directly engaged, the aim is to target the thinking of decision-makers within the backing entity through the S-APT operator, by using that operator to pass on misleading information (see the forthcoming disinformation model).

We develop the Advanced Persistent Threat Operational Line (APTOL) framework to interpret the strategic (goal-directed) objectives of S-APT operators, evaluate the maturity of their operations, disrupt these operations in effective ways, and even to target the operators and their chains of command for certain types of information attack. Although we apply this framework to S-APT operators, the framework itself can be used to interpret the objectives of any APT operation.

To develop this framework, we synthesize principles from military doctrine with findings reported primarily in peer-reviewed literature. This operational framework links APT tactics together in the context of a coherent operation that is carried out to support strategic objectives (see figure 2). The framework highlights the inherent vulnerabilities of the operation while also identifying interdependency between tactical milestones. There are eight core tactical milestones that every APT operation involves: 1) the selection of the target organization whose network will be exploited; 2) vulnerability assessment prior to tactical targeting for penetration; 3) maneuver/fires to achieve penetration; 4) the unfolding process of infiltration; steps taken to assure 5) sustainability and 6) mobility; and 7) the focused intelligence, surveillance and reconnaissance required for 8) action upon specific data targets. These are eight actions that must be successfully performed for the mission to succeed.

In cases of sabotage, milestones may be accomplished outside of the system that will ultimately be targeted for sabotage. In the historical cases mentioned above (Stuxnet, Shamoon 2, and Crash Override), malware is deployed which achieves milestones 3 forward by design, and situation awareness is a factor only during development of the malware because there is no human involvement after its deployment. This is to say that the developer(s) must have awareness regarding the situation as defined by design requirements for assuring operational security and task execution, given known characteristics of the target environment.

Although many tasks within the operation can be automated by programming them into malware, the performance of these tasks must still be monitored to some extent by a human operator to ensure that strategic objectives are being met. Furthermore, any human decision-making within the context of an operational environment requires situation awareness, i.e. an accurate understanding of what is going on within that situation. The operator must be situation-aware in the interests of both attack and defense, i.e. pursuing victory while also avoiding defeat. This can be expressed in terms of situation awareness for task execution (i.e. awareness regarding what the attacker must do to achieve tactical milestones); and situation awareness for operational security (i.e. awareness regarding what the attacker must do to avoid detection). Intelligence is required for command and control over the operation and can be synonymous with situation awareness within the context of the operation itself.

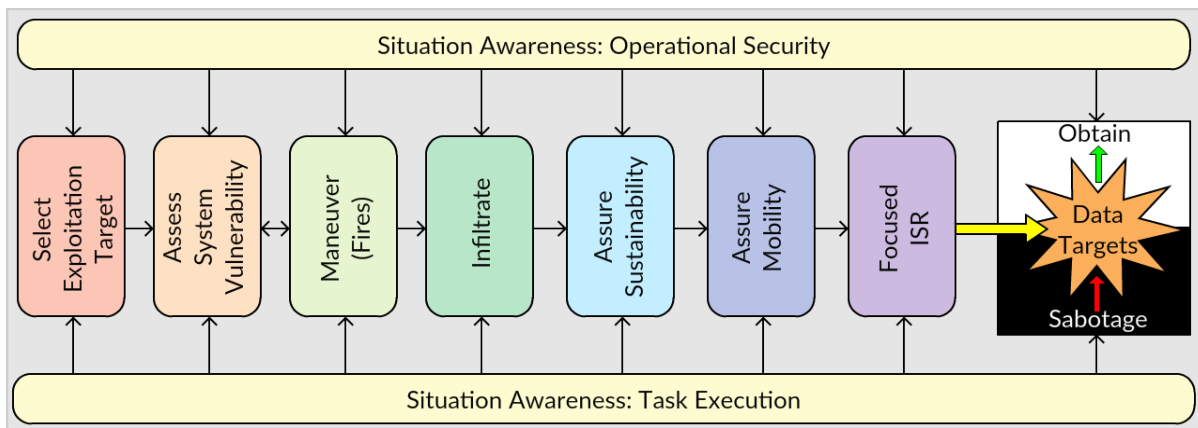

**Figure 2: The APTOL**

Proceeding from left to right, the first step in the APTOL is the selection of a target system/network for intrusion/exploitation. The target is selected based on prior or current intelligence, surveillance and reconnaissance (ISR) on systems belonging to people and organizations that are considered relevant to the attacker for whatever reason; selection criteria relate to the expected value of the target vis-à-vis the attacking entity's strategic objectives (i.e. gaining wealth or achieving competitive advantage). Next, the vulnerability of the target is assessed using either invasive or non-invasive means. As attack can be employed as an invasive means of assessment, maneuver (movement combined with fires) may occur prior to, simultaneously with, or after the making of a vulnerability assessment.

Where assessment has not already culminated in access, the APT operator then applies fires (one or more separate technical capabilities) to induce an effect on the target that will facilitate intrusion (e.g. a software feature is exploited to reveal a list of user credentials, which are then used to access the system). The operator then infiltrates the network, which entails movement between nodes; general ISR; maintaining situation awareness; disabling defenses through application of fires (as necessary); and employing measures in the interest of evading detection.

To ensure sustainability of operations, the operator's next priority is to develop redundant access (securing credentials, introducing backdoors, etc.) to the target system(s). After this has been assured, the operator can then work to establish the level of access/mobility required for accomplishing the goals of the operation (i.e. by achieving access to all network segments of interest; creating secure avenues for data exfiltration; etc.). Reconnaissance and surveillance is then carried out to obtain or sabotage[5]. To avoid detection and to operate effectively in support of strategic objectives, the operator must maintain situation awareness in operational security and task execution, respectively, throughout the operation.

## 7.0 Situating APT Tactics in a Strategically Motivated Operation

In this section we explain how S-APTs execute their strategically motivated operations using TTPs. We engage in a rich discussion of S-APT TTPs while also placing these TTPs in their proper context as elements of human-driven, strategically motivated operations. We have included some additional information from non-peer reviewed sources to support this core material where we feel it can lend more insight for practitioners; included in this category are specific details on TTPs applied during the Triton/Trisis Schneider Electric operation conducted in August 2017.

The Triton/Trisis operation targeted a petrochemical plant in Saudi Arabia (still unidentified at the time of writing), with the apparent goal of sabotaging the plant's operations in a way that could have resulted in people being seriously harmed or killed (Perlroth and Krauss 2018). The attack involved intricate original malware (referred to sometimes as "Triton," sometimes as "Trisis") which was designed to enable exploitation of the Triconex Safety Instrumented System (SIS), a logic controller "primarily used to manage physical equipment in nuclear power plants, oil and gas production facilities and paper mills" (Bing, 2018). Luckily, due to a minor programming error that inadvertently triggered an emergency shutdown, the operation was both unsuccessful and detected (Newman 2018).

### *Target Selection*

The attackers behind S-APTs tend to go after organizations known or expected to have data of high financial or strategic value (SBIC 2011; Maisey 2014; Wangen 2015; Lemay et al. 2018). This includes any information that can be sold, used to access funds, or otherwise used to make money; as well as any information that can support planning or execution of activities expected to gain competitive advantage or otherwise improve the relative positioning of the obtaining party or backing entity. S-APTs may be prosecuted to serve the strategic interests of criminal enterprises, rival corporations, or foreign governments (SBIC 2011; Tankard 2011; Maisey 2014; Nelson 2014 Jasper 2015). Boundaries between these group types are sometimes indistinct (Smiraus and Jasek 2011; Mansfield-Devine 2014). When attacks are carried out by organized criminals, the strategic objectives are relatively narrow, i.e. always relating either directly or indirectly to financial gain (Scully 2011). When attacks are carried out for corporations or nations, the strategic objectives are relatively broad, i.e. competitive advantage, or improving the entity's status relative to other entities within a national or international political-economic system (Scully 2011; Potts 2012; Mawudor 2013; Jasper 2015).

Organizations are targeted for what they have, i.e. the kinds of information they handle; or what they do, i.e. their roles within larger economic, political, or social systems (SBIC 2011). S-APTs have afflicted broadcast; defense; financial services; oil-and-gas; online gaming; market-services; security; and technology industries, in addition to government offices worldwide (SBIC 2011). In cases where the threat is reported to be "state-backed," attacks have been directed toward: health insurance providers; the media and entertainment industries; technology sector; political parties; and non-government organizations with activity in democracy and human rights; energy sector; and other elements of critical national infrastructure (SBIC 2011; Rice 2014; Maisey 2014; Wangen 2015; Lemay et al 2018). Perhaps the most common targets reported are major telecommunications corporations, government organizations, and corporations directly tied to defense (SBIC 2011; Maisey 2014). Less information is available on such attacks against financial services, energy/utilities, transportation and shipping, though they are likely to have been the subjects of "preparatory and ongoing compromises" (Maisey 2014, 5). The majority of attacks on industrial supervisory control and data acquisition (SCADA) systems have been reported to be S-APTs (Kim et al. 2014).

In some cases, the only targeting criteria applied by S-APT operators may be the industry sector that the organization belongs to (Kim et al. 2014; Caldwell 2013; Auty 2015). The mission goals of S-APTs can be the obtainment of specific kinds of information or the sabotage of operations (Hutchins et al 2011; SBIC 2011; Brewer 2014; Arquilla 2015; Jasper 2015), to serve strategic objectives of enrichment (material gain) or competitive advantage—whether at a corporate or national level (Scully 2011; Wangen 2015; Lemay et al. 2018). Information of interest includes information pertaining to individuals working in or connected to the target organization; the design/structure and composition of the organization's information systems; and the design and functionality of business processes (SBIC 2011). Sometimes an organization is targeted only as an entry point into another organization with which the initial organization has some business relationship (Scully 2011; Caldwell 2013; Caldwell 2015).

Information/intelligence about targets is often collected via "open" or publicly available sources (Sood and Enbody 2013) such as industry websites, conference proceedings, publicly available reports (Hutchins et al. 2011), press releases and employment advertisements (Julisch 2013), online social networking sites (Cutler 2010; Julisch 2013; Caldwell 2013; Sood and Enbody 2013), or information found in organizational trash receptacles (Krombholz et al. 2015). Intelligence collection on the target organization may persist for some time prior to launching the actual attack (Sood and Enbody 2013).

---
[5] Where data theft involves the complete removal of target data from a system, it can constitute both obtainment and sabotage.

The target of the Triton/Trisis operation was a high-profile target, presumably selected for its role in the Saudi economy (Perlroth and Krauss 2018). Investigators concluded that the operation was probably state-backed because it would have been expensive to plan and execute despite no immediate financial gain being involved (Perlroth and Krauss 2018; Bing 2018). Some have connected the attack to Iran (Kovacs 2018; Perlroth and Krauss 2018). Planning for the operation is estimated to have taken over a year (McMillan 2018).

### *Assessing System Vulnerability*
Once a network has been selected as a target, it is assessed for vulnerability and the mission becomes oriented toward gaining access (Caglayan et al. 2012; Brewer 2014). Vulnerability assessment may require less intensive efforts or more intensive efforts, depending on what is already known about the target. Assessment may focus on discovering or verifying the presence of technological vulnerabilities; profiling individuals to varying degrees (i.e. in the interest of crafting deceptions that will enable operators to introduce malware or elicit credentials through social engineering); or both (Smiraus and Jasek 2011; Brewer 2014).

Assessing technological vulnerabilities of targets can be done intrusively ("typical instruments used by pen-testers to discover network services, to identify vulnerabilities and to actively try to exploit these vulnerabilities") or non-intrusively ("information originating from system descriptions, human operators, running services, but also soft scanning techniques, including slow-rate scanning, identification of missed update notifications, registry entries, etc.") to identify one or more points of vulnerability that can be leveraged to gain access (Genge et al. 2015, 14 and 13). Typically, non-intrusive targeting is not detected by the target (Sood and Enbody 2013). Where assessment is intrusive, identification and exploitation of vulnerability may occur simultaneously; where assessment is non-intrusive, the vulnerability is discovered first and an attack is then customized to exploit it (Sood and Enbody 2013; Brewer 2014). In some cases, a known vulnerability is simply targeted with the expectation that it is probably present in the targeted system (Caldwell 2013).

Vulnerability assessment may also focus on human fallibility in the interest of social engineering. Here, information about people and their roles is gathered to support an indirect approach in which someone else is deceived into exploiting a technological point of vulnerability, e.g. by opening a malware-embedded email attachment or following a hyperlink to a malware-infected site (Smiraus and Jasek 2011; Caldwell 2013). In other cases, an indirect approach need not leverage technological vulnerability at all, as sensitive information such as usernames and passwords might simply be elicited through social engineering tactics (Tankard 2011; Krombholz et al. 2015). Depending on the situation, a combination of direct and indirect approaches may be used to exploit technological vulnerabilities (Steer 2014; Wangen 2015; Lemay et al. 2018). Some intrusions have been facilitated through the distribution of hardware with built-in backdoors— "malicious firmware" (Sood and Enbody 2013).

S-APT operators are often well prepared prior to undertaking an attack and exploit operation, and this preparation may include reconstructing a target organization's defenses in a controlled environment to practice defeating them (Bradbury 2010; Sood and Enbody 2013; Caldwell 2015). The preparations for attacks may take many weeks or months (Smith 2013). In some cases, malware is custom written originally, or novel approaches devised, especially for use on the target organization's systems (Scully 2011; Smiraus and Jasek 2011; Brewer 2014). Where accomplishing a breach will rely on spear phishing, the design and content of the email and attachment that will deliver the malware payload must be both convincing and provocative in design to invite inspection (Caldwell 2013; Brewer 2014). Arriving at such a design may require gathering some intelligence on another organization to impersonate it in communications (Hutchins et al. 2011); a degree of research into the target individual's personal or professional life; or a degree of subject matter knowledge relating to the target's personal or professional interests (Caldwell 2013).

Investigators surmise that the attackers behind the Triton/Trisis operation must have had their own copy of the Triconex safety system, which they then reverse engineered to identify vulnerability and plan exploitation (Newman 2018; Perlroth and Krauss 2018). It is a feature of the system that it reprogrammable when a physical switch on the workstation device is set to "program," and the switch had been left in this position at the targeted plant (McMillan 2018). Schneider Electric stated that programming could then be altered via "the TriStation terminal or another machine connected to that safety network" (Hand 2018). Planning the sabotage itself would have required knowledge of the layout of the pressure regulation system and how an explosion could be induced by altering the functionality of that system once the safety system was effectively disabled (Perloth and Krauss 2018; Bing 2018). To some extent this intelligence could have been collected through Triton/Trisis, as it is designed to conduct "system analysis and reconnaissance" (Newman 2018).

### *Maneuvering (fires)*
The operator may attempt to attack public facing systems, internal network devices or client machines (PCs, mobile phones, tablets etc; Li et al 2016) using a direct exploit (i.e. by introducing code), using a known or discovered vulnerability (Bann et al. 2015); an indirect exploit (having someone else—usually unwittingly—introduce code that exploits a known or discovered vulnerability); or some combination of the two to enter a system or network (Steer 2014; Bann et al. 2015; Li et al 2016). Commonly employed approaches to direct exploitation have included "pass-the-hash" attacks, SQL injection attacks (Sood and Enbody 2013; Bann et al. 2015); cross-site scripting, "use after free" attacks, and DNS cache-poisoning (Sood end Enbody 2013).

Common approaches to indirect exploitation include introducing malware via an infected website (Bradbury 2010; Smith 2013; Auty 2015); email attachment, e.g. a malware-embedded PDF or Office document (Hutchins et al 2011; Caldwell 2013; Smith 2013; Nissim et al 2015); or removable media, e.g. a USB drive (Hutchins et al. 2011; Julisch 2013; Jasper 2015). More recently the security press and CERT community have reported activity using indirect attacks to deliver payloads by different mechanisms, including: in June 2017 via modified software updates (Tung 2017); in September 2017 via modified signed software (Goodin 2017); and in November 2017 via the Microsoft Office Dynamic Data Exchange protocol (Palmer 2017; SingCERT 2017). Reported blended approaches include: using fake websites to obtain credentials (Li et al 2016); using fake wireless networks to launch attacks or capture sensitive information (Li et al 2016; Kolias et al 2016); or connecting a malware containing device (Brewer 2014; Kim et al. 2014; Krombholz et al. 2015; Wangen 2015; Lemay et al. 2018).

In the first case, a legitimate site may be compromised (e.g. a "watering hole" or "drive-by download" attack) and users attempting to access it are redirected to an illegitimate, malware infected site (Julisch 2013; Steer 2014; Auty 2015; Krombholz et al. 2015). In the second case, a legitimate-looking email is sent to the recipient along with an attached file designed to exploit vulnerability at the file-based-communications level (Caldwell 2013; Nissim et al. 2015), i.e. in virtually any application associated with opening and reading a file of a particular type (Nissim et al. 2015). In the third case: a legitimate-looking email is sent to the recipient encouraging them to click a link to redirect them to a website designed to capture their credentials when the user attempts to login (Li et al 2016), this information can then be used to launch further attacks; or a wireless network is impersonated, encouraging users to connect in order to probe for and attack discovered vulnerabilities and capture traffic to reveal sensitive information (Li et al 2016; Kolias et al 2016); or a device is connected to a system by an insider who either intentionally or unintentionally transmits malware to the system in this fashion (Brewer 2014; Kim et al. 2014; Krombholz et al. 2015; Wangen 2015; Lemay et al. 2018).

Indirect approaches can enable the attacker to completely circumvent the technological perimeter defenses that most organizations rely on for security (Scully 2011; Potts 2012; Caldwell 2013; Sood and Enbody 2013). More reported S-APTs have been initiated through spear phishing than through any other approach (Brewer 2014; Caldwell 2013; Nissim et al. 2015). An infected file may be of the type indicated by its extension (with malicious code embedded within it) or the extension may itself be deceptive (Brewer 2014). When the attachment is opened, it (typically covertly) activates malware embedded in the file (Hutchins 2011; Brewer 2014; Caldwell 2015; Nissim et al. 2015). Depending on its complexity, the activated malware may begin to automatically execute a plethora of different functions that carry on throughout the life of the mission (Munro 2012; Wangen 2015; Lemay et al. 2018).

Triton/Trisis was introduced "remotely" (Perloth and Krauss 2018), presumably through a compromised computer connected to the safety network (cf. Hand 2018, above). The malware was designed to exploit a specific firmware vulnerability present in the version of the Triconex safety system that was in use at the targeted plant (Newman 2018).

### *Infiltration*
Through the exploit, access is achieved. The S-APT operator (or preconfigured malware already introduced into the target system) then conducts "mapping" or "enumeration" to "fingerprint" the target network; a reconnaissance tactic to identify open ports, access points, addresses, firewall or intrusion detection or prevention system characteristics, the currency of active systems, running software applications or services, virtual hosts or platforms, or storage infrastructure that can then be exploited (Smiraus and Jasek 2011; Tankard 2011; Sood and Enbody 2013; Caldwell 2015). The network map is then used to plan courses of action, devise tactics, and select tools appropriate for the mission (Tankard 2011; Julisch 2013; Caldwell 2015). In cases where penetration is achieved via a malware-embedded email attachment, there may be little knowledge of network architecture prior to penetration, and the attacker may need to discover where within the target network access has been gained (Smiraus and Jasek 2011), and whether the compromised system is mapped to the network drive (Tankard 2011). Mapping is often accomplished using normal commands that do not stand out as anomalous events (Moon et al. 2014).

Once access has been achieved and the mission is underway, the S-APT operator's objective is to remain undetected and operational for as long as necessary or desired (Tankard 2011; Maisey 2014). S-APTs can sometimes involve the maintenance of persistent, undetected access for years (Scully 2011; Auty 2015; Caldwell 2015). S-APTs are often marked by use of original/custom programming code (a "zero-day exploit") or attack patterns that have been designed specifically to flout automated detection (SBIC 2011; Brewer 2014; Mansfield-Devine 2014). The S-APT often attacks defense systems in the interest of deceiving or disabling them (SBIC 2011; Brewer 2014; Friedberg et al. 2015; Wangen 2015; Lemay et al. 2018), or to create diversions that distract from key tactical movements (SBIC 2011). This can include the deletion of data from logs to erase forensic evidence (Smiraus and Jasek 2011; Auty 2015; Caldwell 2015) or launching a DDoS attack that serves as a diversion "...whilst a vulnerability is exploited elsewhere to ingress malware or hack pre-planned entry" (Gold 2014). Because conspicuously missing data can serve as a reliable indicator that intrusion has occurred (Caldwell 2015), incriminating log data is sometimes replaced with crafted disinformation (Maisey 2014; Wangen 2015; Lemay et al. 2018).

Overtly anomalous behavior must be avoided in the interest of evading detection, so the S-APT operator must ensure that mission tasks are carried out in subtle and inconspicuous ways (Caldwell 2015). Once the S-APT operator is established within the target network, his or her activities are generally difficult to detect because they are typically executed via standard or otherwise inconspicuous channels, using normal communication protocols, for example FTP, HTTP/S, SMTP, DNS or NTP (Tankard 2014; Bann et al. 2015; Kovacs 2018; Mitre 2018). Both command and control and the movement of data commonly happen over inconstant, low-density signaling that is difficult to identify as a persistent anomaly (SBIC 2011). Malware is often stored or deployed in locations where it is unlikely to be noticed, e.g. "servers, routers, printers and wireless access points" (Smith 2013, 19).

Conduct of an S-APT operation generally requires software tools. In some cases the initial malware infection includes all of the functionality desired for a particular attack (Raiu 2012; Sood and Enbody 2013; Wangen 2015; Lemay et al. 2018). In other cases, a rudimentary malware program is designed to build upon itself by downloading additional components slowly over time (Sood and Enbody 2013; Brewer 2014; Auty 2015). Sometimes the malware introduced during the breach is not enough to accomplish the goals of the mission and the S-APT operator must download additional malware tools to one of the compromised systems within the target network as the mission develops (Smiraus and Jasek 2011; Steer 2014).

Tools used in APTs include "key loggers, Trojan backdoors, password crackers, and file grabbers" (Steer 2014). Hutchins et al (2011) argue that the kinds of malware used in S-APTs "especially requires manual interaction" (116). Offensive tactics must often be adapted to defeat or circumvent new or changing tactics employed by defenders (Kim et al. 2014). Commonly, malware is dynamically reconfigured or recalibrated in the interest of defeating or evading defensive measures or countermeasures (SBIC 2011; Munro 2012). It is common practice to "upgrade" and rewrite some elements of code while retaining the malware's core program, to evade detection while maintaining the desired functionality (Auty 2015). When malware is designed to automatically seek out certain types of data, the search terms guiding this behavior may be modified to refine, expand, or otherwise change targeting priorities (Smiraus and Jasek 2011).

Stealth is often supported by rootkit-enabled software that maximizes control while dynamically disguising its own presence or the presence of other software tools (Thomson 2011). Encryption-based obfuscation techniques used to hide or disguise malware include code that is designed to randomly or systematically change itself or its decryption routines, or the hiding of executable code within a compressed or encrypted file format so any signature associated with its contents is not detectable by antivirus software (Brewer 2014; Sood and Enbody 2013; NetworkWorld Asia 2014; Alazab 2015; Caldwell 2015; Nissim et al. 2015).

Specific toolkit types include packers that compress malware files to reduce their size; crypters that encrypt content to hide signatures; code protectors that are resistant to debugging or examination within virtual environments; or packagers that misrepresent malware as—or embed it within—a legitimate program (Sood and Enbody 2013). Other techniques used to evade signature detection include the use of intentionally complex or otherwise unusually formatted command code; the breaking of code up into multiple files that appear to be innocuous when examined separately; or writing code in such a way that it is structurally interrupted but functionally coherent, e.g. integrating white space or additional code that is ultimately filtered out, leaving only a string of now uninterrupted malicious code to run (Nissim et al. 2015).

An S-APT can involve long periods without any activity, during which implanted malware simply lays dormant (Brewer 2014; Auty 2015; Caldwell 2015). In other cases, the operator posing the S-APT may have abandoned the use of malware completely and is only using legitimate components of the operating system, e.g. remote desktop access functionality and credentials with administrator-level privileges (Smith 2013; Caldwell 2016). Here, the threat is persistent, but "advanced" only insofar as it is not readily or easily detected by standard security technologies, i.e. it is not reliant on ingenious software design or artful decision-making by an expert, but can still circumvent standard defenses (Raiu 2012; NetworkWorld Asia 2014; Richardson 2015). Such malware-less "living off the land" (Richardson 2015) attacks can be perpetrated almost entirely using built-in functionalities, for example Microsoft NT LAN Manager (NetworkWorld Asia 2014; Preempt 2017) or Windows Management Instrumentation (Richardson 2015; Microsoft 2017). In these cases, the S-APT operator is effectively moving as an insider and may only be exposed through event correlation and behavioral analytics (Caldwell 2015; Friedberg et al. 2015; Richardson 2015; Warren 2015).

Weak approaches to log analysis/anomaly detection decrease the chances of detecting the presence of an S-APT (Cutler 2010; Tankard 2011; SBIC 2011; Friedberg et al. 2015). Friedberg et al. (2015) point to the importance of "mining…typically unnoticed relationships between different applications and components of a network" to model network behavior patterns accurately enough to identify subtle correlations that can be used to more reliably distinguish between normal and anomalous activity (36). Detecting and countering S-APTs requires proactively collecting and analyzing intelligence from internal sources such as logs, as well as from external sources such as government bodies or other organizations (SBIC 2011; Julisch 2013; Webb et al. 2014; Webb et al. 2016). Although technology is required to collect and pre-process data relating to technological indications across logs (Tankard 2011; Maisey 2014; Warren 2015), Maisey (2014), Webb et al. (2014) and Webb et al. (2016) have stressed the importance of having specialized human analysts interpret the real significance of this data.

Upon exploitation of the vulnerability by Triton/Trisis, privileges were escalated and a remote access trojan (RAT) was injected into memory (Osborne 2018), enabling system control by the off-site S-APT operator (Kovacs 2018). The malware's built-in functionality enabled it to "scan and map the industrial control system and conduct reconnaissance" (Osborne 2018).

### *Assuring Sustainability*

The sustainment of S-APT operations is to some extent dependent on the maintenance of infrastructure external to the target network. Prior to launching an attack, "command and control" servers—which will be used during the attack to focus efforts, channel traffic and operate malware—must be established and available (Tankard 2011 and 2014; Caglayan et al. 2012). Command and control servers are typically computers that were compromised during previous missions (Bradbury 2010; Raiu 2012). Massive offenses have led to the compromise of systems around the globe to create "botnets" that can put tremendous computing power behind S-APT missions (Bradbury 2010; Sood and Enbody 2013; Caldwell 2015). A command and control server may be used to control multiple botnets consisting of numerous compromised systems that are then used to carry out various functions entailed in an S-APT attack, i.e. "one botnet for social engineering (spear phishing), another to infect machines and proxies, and still others to exfiltrate the data" (Caglayan et al. 2012). The command and control server may be in the same country as the target or embedded within infrastructure as a service (IaaS) to make communications with its IP address less conspicuous (Tankard 2014; Wangen 2015; Lemay et al. 2018).

Assuring sustainability for an operation is largely a matter of maintaining operator access to the network. In many instances, some assurance of sustainability is accomplished with the initial attack. Penetration via spear phishing, for example, may introduce a malware combination that enables immediate access by exploiting some software vulnerability present in the target system, while also enabling persistent access via the establishment of a clandestine communications channel (Cutler 2010; Tankard 2011; Hutchins et al 2011; Brewer 2014). This channel, created by a RAT (remote access Trojan—or toolkit) may be designed to install itself onto a host and run as a service (Brewer 2014; Caldwell 2015), enabling constant access to the network by the S-APT operator. The RAT is configured to initiate an outbound ("reverse") connection (Chang et al. 2016) between the victim system and the attacker's command and control server, and typically communicates back to the operator with encrypted traffic sent over port TCP 443 (HTTPS) to the command and control server, to appear as normal outbound web traffic (Brewer 2014; Moon et al. 2014). The RAT calls back to the operator at designated intervals to indicate that the backdoor remains open and enabling communications between itself and the command and control server to appear as trusted (Smiraus and Jasek 2011; Steer 2014).

Researchers from ESET report a technique that is particularly difficult to detect and which is thought to have been used by APT28 to assure sustainability. This involved using an UEFI root kit dubbed Lojax to attack misconfigured or vulnerable computers to obtain persistence that is resistant to replacement of storage devices or reinstallation of operating systems (ESET 2018). ESET report that the Lojax root kit is installed by a combination of custom and freely available tools which are most likely delivered to the target system by an initial infiltration technique (ESET 2018). The root kit writes and installs a maliciously modified version of the small agent component for Lojack anti-theft software to the infected computer's storage device each time the system firmware selects the boot device at startup, while the operating system is loading, thus ensuring persistence (ESET 2018). Lojax small agent acts as a service and is thought to be able to connect to command and control servers to download additional tools as required (ESET 2018). ESET propose that the attack may go undetected because the legitimate Lojack software is often whitelisted by antivirus software, making it more difficult to detect the modified Lojax small agent (ESET 2018). ESET note from their investigation that the use of the Lojax UEFI root kit was rare and used to target "mostly government entities located in the Balkans as well as Central and Eastern Europe" (EST 2018).

As a single back door would constitute a single point of failure for the whole operation, S-APT operators generally take additional steps to ensure persistent and redundant access, i.e. via implantation of multiple backdoors/remote access tools (Smiraus and Jasek 2011; Tankard 2011; Brewer 2014; Caldwell 2015), or by securing multiple user credentials (SBIC 2011; Krombholz et al. 2015). This redundant accessibility naturally supports continued access even if one or more backdoors or compromised accounts are discovered and eliminated (Brewer 2014; Caldwell 2015).

Credential theft is the most common way to maintain access to a network (Brewer 2014), and a higher level of assurance can be afforded through the theft of multiple user credentials (Krombholz et al. 2015). Malware is sometimes programmed to automatically "harvest" credentials and escalate privileges on them to gain administrative rights (Thomson 2011; Chang et al. 2016). Credential theft helps to assure redundant access after the initial compromise (Smith 2013; Brewer 2014). Often a lower-privileged user's credentials are the first to be stolen and privileges are then escalated by stealing additional, higher-privileged credentials once access has been gained (Caldwell 2015), or by exploiting software vulnerabilities to assign higher privileges to credentials already in possession (Chang et al. 2016).

Stolen credentials that enable the S-APT operator to simply log on and navigate the network unnoticed may be all that is required to achieve mission objectives (Smith 2013). If sufficient privileges can be obtained by the S-APT operator (e.g. captured through manual or automated information gathering), the use of malware may no longer be necessary (Brewer 2014; Caldwell 2015). In some cases, the attacker may create a new user account separate from the organization's extant accounts (Caldwell 2015), though this risks detection where accounts are monitored for such changes.

A "configuration mistake" (Bing 2018) written into Triton/Trisis led to application code failing a validation check between redundant processing units, which in turn initiated a safe shutdown mode by the safety instrumented system controller (Hand 2017). This event led to a forensic investigation which uncovered an executable file ("trilog.exe")—disguised to look as though it came from Schneider Electric but, upon inspection, created by someone outside of the company—with an attached zip file ("Library.zip") found to contain Triton/Trisis, effectively terminating the S-APT operation (Bing 2018). If the configuration mistake had not been made, no anomaly would have been detected, the shutdown would not have occurred, a forensic investigation would not have been launched, and the operation might have been sustained "for years" (Kling, in Hand 2017).

### *Assuring Mobility*

Where interest is general, the S-APT operator may seek to maintain an expansive presence within the network by distributing malware across any systems of interest (Julisch 2013; Brewer 2014). However, S-APT operators may also tightly limit their attacks on systems within the network to only those required for accessing very specific data, in the interest of avoiding detection (Auty 2015; Caldwell 2016). Gaining access to additional systems following the initial breach may involve further enumeration/mapping; new iterations of preparation and targeting; further escalation of privileges on available credentials; or the importation of additional software/tools (Smiraus and Jasek 2011; Tankard 2011). Some malware used by S-APT operators is configured to seek out and compromise other systems automatically (Sood and Enbody 2013).

Essentially, assuring mobility is synonymous with maintaining access to all network nodes necessary for conduct of the S-APT operation. To a large extent the ability to access different nodes within a network is determined by network architecture—how directly interlinked (as opposed to segmented by bridging devices or firewalls) these nodes are (SBIC 2011; Caldwell 2015). Red-black network segmentation—or ensuring that systems holding the most sensitive information are not connected to the internet in any way—minimizes access to high-value data targets (Caldwell 2013). It has also been argued that cloud computing can be used to make target reconnaissance more difficult for attackers, due to the segmented and distributed nature of resources stored in the cloud (Dorey and Leite 2011).

In the case of the petrochemical plant targeted in the Triton/Trisis Schneider Electric operation, additional measures to assure mobility were not required because, "against best practices," the plant's safety instrumented system was connected to its operations network, meaning that both could be manipulated via Triton/Trisis once the software had been installed (Hand 2017).

### *Focused Intelligence Surveillance and Reconnaissance*

The maintaining of a constant presence generally enables in-depth study of the targeted organization's characteristics and functionality while also decreasing the chances of missing any small-window opportunities to advance the mission (Bradbury 2010; Cutler 2010). In many cases S-APTs use malware that has been designed to systematically compromise systems within the network and collect certain types of information (Smiraus and Jasek 2011; Dorey and Leite 2011; Steer 2014; Wangen 2015; Lemay et al. 2018). The S-APT operator ultimately targets systems to directly access data, but both targeting and attack can focus on people known or expected to have access to—or possession of—the target data (Tankard 2011). For example, a CEO is a high-value target because he or she is likely to possess or access sensitive business information that might not be found elsewhere in the network (Cutler 2010; Tankard 2011).

There are two types of information being collected. The first type is information to support the conduct of the mission, e.g. network layout and composition; credentials; network traffic flows at certain times, etc. (Brewer 2014; Moon et al. 2014). Diagnostic information pertaining to the bounds of normal activity within the network is particularly important, as malware and operator behavior must occur well within these bounds to avoid detection as anomalous patterns (Brewer 2014; Caldwell 2015). The second type of information is information directly related to the operational and strategic objectives of the mission (Smiraus and Jasek 2011; Brewer 2014; Wangen 2015; Lemay et al. 2018). Whereas priorities relating to general target types are typically established prior to mission execution, the targeting of specific pieces of data may be approached generally (i.e. captured en masse) or in a more targeted fashion (Smith 2013).

The data targeted for obtainment or sabotage may be identified through operator-conducted searches or it may be automatically identified by malware according to rules, e.g. specified file types, keywords, metadata, etc. (Smith 2013; Wangen 2015; Lemay et al. 2018). Past targets for obtainment (information types that support the strategic objectives of the attacking team or backing entity—as opposed to information types that support conduct of the S-APT mission itself) have included "technology blueprints, proprietary manufacturing processes, business plans, pricing documents, partnership agreements, emails and contact lists," for attacks likely to be state-sponsored; and "names, addresses, email addresses, birthdates, usernames, passwords, security questions and possibly credit card details" for an attack that was likely perpetrated by criminals in the interest of fraud or financial gain (Ring 2013, 5). It has been reported in the press that the Equifax breach announced in September 2017 was the action of a state sponsored APT, using intelligence gained from initial infiltration to escalate to a more advanced team, redirecting the attack towards high value personal financial information and obtaining intelligence on specific individuals for use in future missions (Riley et al 2017). Effectively defending assets from S-APTs requires understanding what data targets S-APT operators are after (SBIC 2011).

Uncovering indications of targeting behavior—e.g. by scrutinizing histories of direct access to database servers (to include any instances of disabled logging) or identifying that large amounts of data have been copied (Friedberg et al. 2015)—is effectively the defender's last chance to prevent (or at least diminish) the success of an S-APT mission.

As cited previously, Triton/Trisis facilitated system monitoring (Osbourne 2018). This would have provided valuable intelligence regarding characteristics of both the safety instrumented and industrial control systems in use at the targeted plant (Kovacs 2018), which could then be applied toward inducing an equipment failure at the facility (Perlroth and Krauss 2018). Because the operation was terminated with the discovery and removal of the remote access trojan while the plant was shut down, the specific effects the S-APT operator intended to induce are unknown (Newman 2018). That said, Perlroth and Krauss (2018) report that "All of the investigators believe the attack was most likely intended to cause an explosion that would have killed people."

### *Action on Data Targets*

When a system is positively discovered to hold or handle data of interest to the S-APT operator, special malware may then be imported especially to that system to capture or act on the target data (Thomson 2011). Data to be obtained may be accessed and copied, captured through general surveillance measures, or moved (Munro 2012; Maisey 2014; Friedberg et al. 2015). Some malware is designed to collect certain types of data automatically, e.g. by "logging keystrokes, capturing screenshots, audio, webcam photos, geolocation data, copying files to a remote location or a special USB device, (or by) hijacking clipboard and other data" (Mansfield-Devine 2014, 15). Sabotage can occur through the manipulation or removal of data (Hutchins et al. 2011), or through the introduction of spurious data or information (Min and Varadharajan 2014; Jasper 2015) and can be conducted subtly over long periods of time (Min and Varadharajan 2014; Jasper 2015; Wangen 2015; Lemay et al. 2018). Outcomes of sabotage can include the actual physical destruction of equipment when it has been reprogrammed to systematically malfunction (Jin 2015; Wangen 2015; Lemay et al. 2018).

Where information obtainment is the goal, data is exfiltrated out of one or more systems in a network. Exfiltration may involve moving large amounts of data out of the victim network at one time, but S-APTs usually involve the exfiltration of data in small, inconspicuous amounts over a long period of time (Kim et al. 2014; Tankard 2014; Caldwell 2015). Exfiltration may be timed to occur during high traffic times of day to draw less attention (Tankard 2014). The server to which data is exfiltrated may also be the command and control server for the operation, though this is usually not the case (Tankard 2014). In some attacks collection and exfiltration have occurred immediately and automatically upon the activation of malware such as a Trojan (Hutchins et al. 2011). If exfiltration activity has been identified, defenders may choose to keep the communication channel open in the interest of intelligence collection, to inform current and future defenses (SBIC 2011; Auty 2015), or in the interest of deception (SBIC 2011; Maisey 2014).

Prior to exfiltration, data is typically hidden in compressed (RAR) files (possibly disguised as another type of file through steganography) that insulate the data content of these files from security and file management software, and then staged for export on a host (Alazab 2015; Brewer 2014; Chang et al. 2016). Files may also be broken up into small groupings of seemingly dissociated data to render them unrecognizable (Moon et al. 2014). The "staging server" may itself be located outside of the targeted organization's network (Chang et al. 2016). Data is generally exfiltrated through a series of compromised computers to obscure the ultimate destination (SBIC 2011; Tankard 2014). Established exfiltration paths may be altered by the operator in the interest of streamlining the process or decreasing chances of detection (Smiraus and Jasek 2011). Numerous systems across numerous countries may be involved (Bradbury 2010; Wangen 2015; Lemay et al. 2018). A system in a country may be chosen specifically to inspire misattribution, i.e. to make it appear as though the attack originated there when it in fact did not (SBIC 2011; Auty 2015). As Auty (2015) writes:

> *...There is still an element of naivety which is that the host country of the IP addresses that are seen to be conducting the attack must be that of the attackers. The truth is that the IP addresses carrying out the attack may just be the last in a long chain of connections (15).*

Finally, conscientious or otherwise rigorous removal of forensic evidence from compromised systems is a common characteristic of S-APTs, especially those believed to be state-sponsored (Scully 2011; Caldwell 2015). In some cases, attackers attempt to remove all evidence of their presence by the time their mission is completed to deny defenders insights into their agendas (Smith 2013; Wangen 2015; Lemay et al. 2018). Malware can be designed to perform evidence erasure—and, ultimately, self-deleting—functions automatically according to rules or when prompted to by an S-APT operator (Moon et al. 2014; Wangen 2015; Lemay et al. 2018). False and misleading evidence may also be planted in the interest of misattribution or to misdirect from the attacking side's actual agenda (Maisey 2014).

A recent report of S-APT activity in the press demonstrates how an S-APT obtained sensitive information from an Australian Defense contractor. Initial entry was gained through exploiting a vulnerability in a public facing helpdesk system, followed by: reconnaissance of the network; dumping of credentials to obtain and maintain privileged access and enable lateral movement; discovery of data of interest; and exfiltration via compressed archives labelled as PNG image files that were hosted on a public facing webserver in the targets network to enable remote download (Crozier & Corner 2017). As noted above, the Triton/Trisis Schneider Electric operation was luckily discovered and interrupted before it could culminate in sabotage via the manipulation industrial control system data.

## 8.0 A Disinformation Model of Counterattack

A truly offensive stance against S-APTs would entail force projection beyond the defended network. Efforts would not only target S-APT methods, but also the operators themselves, their systems, and all the conditions that enable their operation—to include the command and control capacity of any backing entity—across the cyber, cognitive, and physical dimensions of activity associated with information warfare. Legally speaking, such efforts are the purview of nation states that are either officially at war or are intentionally operating within grey areas of international law while conducting intelligence activities. As far as civilians and civilian organizations are concerned, attacking computer networks generally constitutes criminal behavior unless it is being done with the express consent of their owners, e.g. for security testing.

There are, however, lawful ways to passively counterattack adversaries using disinformation. When discussed in previous literature, these methods are typically limited to disinformation about the system environment itself. They involve the use of "twins" or simulations such as "honeypots" (simulated computer targets) or "honeytokens" (simulated target entities other than a computer, such as software) to disorient or otherwise mislead the attacker (see Fraunholz et al. 2018 for a detailed review of such tactics). When used effectively, these technologies can have tactical and operational implications for S-APT operators and even strategic implications for their backing entities. When operators can be tricked into revealing their tactics and their operational goals within a monitored environment it supports the development of countermeasures against them, thereby potentially aborting a current operation while making future operations more difficult to conduct.

Endsley's (1995) situation awareness (SA) theory is a useful theory for explaining how disinformation can be used to adversely impact goal-oriented behavior. In this section we 1) revisit the basic components of SA theory; 2) explain how the theory can be put into opposite terms to provide a theoretical framework useful for understanding the utility of disinformation; 3) explain operational disinformation techniques currently being used by some practitioners; and 4) explain how this disinformation model applies to strategic disinformation techniques where the S-APT operator is a vector rather that the target.

First, we revisit SA theory. The theory holds that SA occurs in three progressive levels. Level 1 SA, perception, occurs when a person senses and recognizes "the status, attributes, and dynamics of relevant elements in the environment" (Endsley 1995, 36). Level 2 SA, comprehension, occurs when a person mentally "forms a holistic picture of the environment, comprehending the significance of objects and events" through "a synthesis of disjointed Level 1 elements" (Endsley 1995, 37). Level 3 SA, projection, occurs when comprehension is thorough enough to project the future of a situation based on "the status and dynamics of the elements and comprehension of the situation" (Endsley 1995, 37). All three levels of SA require the comparison (through autonomic and conscious processes) of incoming sense data and emerging ideations to preexisting reference information available to the decision-maker. Situation awareness is thus the output of a cognitive process carried out to arrive at current knowledge about a situation, and this output becomes the input for goal-oriented decision and action (including conscious inaction) pertaining to that situation (e.g. an environment that an operator must navigate).

Second, we now explain how putting Endsley's theoretical model of SA into opposite terms explains the utility of disinformation (see Figure 3). Starting from the model's left, disinformation is made available within the information environment by the defending side. If, upon comparison to reference information, the disinformation product is perceived legitimate by the attacking side, then misperception of elements composing the actual situation results. When these spurious elements are integrated into an understanding of the overall situation on the attacking side, misapprehension of the situation occurs. Because this misapprehension skews the attacking side's view of relationships between situational elements, it can result in misprediction about how the situation will develop in response to certain interventions. This inaccurate projection by the attacking side in turn results in decision to take a course of action or inaction, which is founded on expectations that do not fully agree with the actual situation (potentially a "bad decision," particularly if failure can be traced back to the decision in retrospect). The effects of the action or inaction can therefore deviate significantly from those anticipated, meaning that the situation can then develop in unintended/undesired ways, possibly to the extent of "mission failure" for the attacking side.

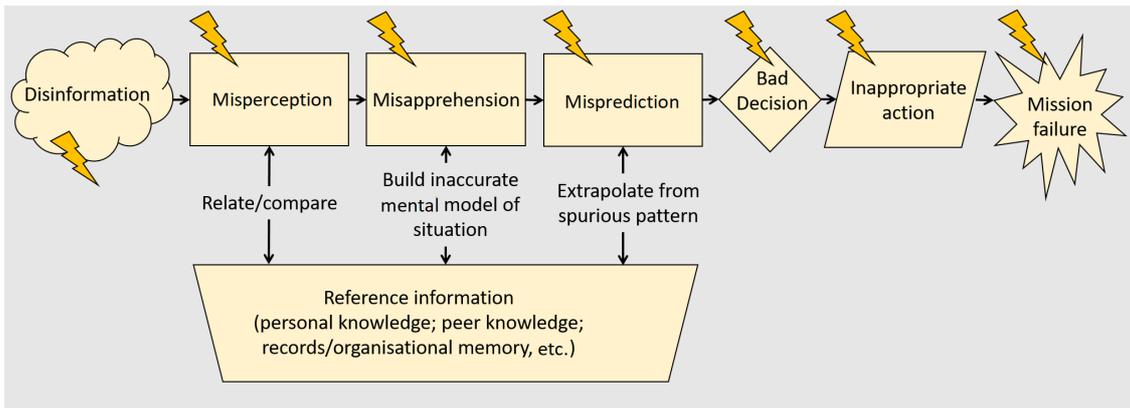

Figure 3: Disinformation to Exploit SA Vulnerability

Third, we now explain how this model applies to current practices in information security. In the case of a honeypot or honeytoken, some asset or complex of assets is perceived real when it is not. For our purposes here, let us consider a complex and convincing honey network that contains honeytokens. The honey network belonging to some organization of interest to the backing entity is selected as a target for exploitation by the S-APT operator. The operator assesses nodes of the illusive network for vulnerability, employs one or more maneuvers to penetrate the specific node(s) and, upon penetration, infiltration of the network begins. If necessary, the operator then employs additional maneuvers to achieve the tactical milestones of assured sustainability and mobility. Focused intelligence, surveillance, and reconnaissance follow to identify specific data targets for obtainment or sabotage, and maneuvers are employed to obtain or modify data. If the deception is intricate and convincing enough, the operator may end up giving away everything from tools, tactics and procedures to ultimate mission goals.

In this example, the operator's situation awareness regarding the execution of mission tasks is entirely spurious because the network and all its apparent attributes are false. More importantly, the operator's situation awareness regarding operational security is entirely spurious because no real operational security has been achieved: his or her maneuvers are being monitored and recorded. As noted above, this can significantly support the defender's defense efforts by showing the defender how the operator does things and, possibly, to what end. These lessons learned can then be used to develop specific countermeasures which defeat the operation and render the maneuvers employed obsolete (unusable in future operations).

Beyond using deception to entrap S-APT operators and learn from them in the interest of developing better countermeasures, some authors have alluded to ways in which S-APTs might be used as vectors for passive counterattack. For example, Almeshekah and Spafford (2016, 32) write of passing on data that can "damage at the adversaries' servers" and Karuna et al. (2018) present a method for generating fake texts that essentially serve as noise to make evaluating stolen intellectual property more difficult. In the former case, the counterattack can have strategic implications for the S-APT's backing entity because it can delay the conduct of future operations; in the latter case the counterattack can have strategic implications because it makes stolen data harder for analysts and decision-makers to understand and learn from. These approaches are more likely to result in temporary setbacks than they are strategic disadvantages for the adversary, however.

This leads us to our fourth goal for this section: explaining how the S-APT operator can be used as vector to target strategic situation awareness and decision-making within the backing entity. In their paper, Almeshekah and Spafford (2016) refer to "the FAREWELL Dossier," which is an excellent historical example of how spies can be used as vectors for disinformation that has significant strategic implications. Here, the United States Central Intelligence Agency (CIA), in coordination with the United States Federal Bureau of Investigation (FBI) (exploiting intelligence provided by a Soviet agent for France codenamed FAREWELL), developed carefully engineered disinformation to disrupt a large scale Soviet industrial espionage program being carried out by agents of the Komitet Gosudarstvennoy Bezopasnosti (KGB) "Line X" team (Weiss 2008; Murry 2004; Kostin et al., 2011; Almeshekah and Spafford 2016). The KGB were fed flawed products and designs with the intention of misdirecting efforts and resources to induce strategic setbacks for the Soviet Union. One of the most spectacular impacts of the CIA disinformation program came from enabling the Line X team to steal altered control system software, which ultimately resulted in the destruction of the trans-Siberian gas pipeline in the biggest "non-nuclear explosion and fire ever seen from space" (Safrie 2004).

To offer an example in this spirit, if a company knows that it would be considered a valuable target for S-APT operations because it is an industry leader, then it can ensure that inaccurate product specifications are routinely stored in a high security network location, fully expecting that this location will eventually be breached by some skilled operator (while genuine specifications are stored completely offline). The S-APT operator, who is looking for information valuable to his or her backers, discovers the dutifully secured product specifications and unwittingly passes them on to analysts or decision-makers within the backing entity. Referring again to Figure 3, when we apply the model to this context it now describes what happens in the minds of these analysts or decision-makers. The "situation" or "environment" concerned is the broader competitive environment within which the backing entity is attempting to advance. Misperceptions now lead to misapprehensions, bad decisions, inappropriate action and mission failure at the strategic level. This mission failure might be measured in terms of misconceptions about supply chain dependency that lead to ineffective plans to control resources or markets (e.g. the specifications deceptively suggest heavy reliance on a particular raw material); the mass-production of defective goods which tarnish the brand (e.g. the specifications misrepresent materials, quantities, or processes involved in production); or time and resources sunk into production equipment or whole facilities that initially work but soon become chronic problems (e.g. because eventual or systematic failure has been built into the programming for equipment routines).

The chain from disinformation to mission failure in Figure 3 is not, of course, a foregone conclusion. It is obviously possible that an attempt at disinformation is unconvincing and is therefore simply excluded from the attacking side's understanding of the situation—meaning that deception does not occur, and the disinformation product does not affect decision-making in the intended way. Such failures can have the undesirable consequence of inspiring the adversary to become more careful. It is also possible that the disinformation is convincing but consumed with unintended consequences, such as the attacker taking a course of action which the defender did not anticipate (e.g. the false product specifications inspire the backing entity to take its own research and development program in a new and better direction that results in market dominance).

So, in all cases—whether the target consumer of the disinformation product is an S-APT operator, an intelligence analyst, or a strategic decision-maker—the product must be designed carefully, using input from subject matter experts who understand how particular kinds of information are useful to others, and with careful consideration of the possible implications. While an information security specialist may be skilled enough to deceive an S-APT operator, deceiving the operator's backing entity will in most cases require collaboration with subject matter experts in other areas. Taking the previous example, if the goal is to lead the backing entity into believing that fake product specifications are legitimate, they need to be good enough to fool an engineer who works in that area. This will almost certainly require input from the organization's R&D team and possibly input from others in business, finance, or another area, depending on the desired effects of the disinformation. Just as a highly skilled S-APT operator can be expected to recognize sophomoric uses of honeypots and honeytokens, an experienced intelligence analyst or decision-maker can be expected to recognize clumsy attempts at strategic disinformation.

Lastly, it should be made clear that while in the FAREWELL example the CIA had obtained a list of technologies that the KGB were known to be pursuing, in many cases the potential value of a type of information can be relatively obvious. In such cases effective disinformation might be made available without requiring any insight into the specific mission objectives of a specific attacker (or even knowledge that an intrusion has taken place).

## 9.0 Discussion

This paper examines the background usage of the term "advanced persistent threat" (APT), provides a new formal definition for the term; identifies the need for a model that reflects the purposeful, human dimension of the threat type known as S-APT; draws on military doctrine to develop the APT operation line model (APTOL), which highlights the inherent vulnerabilities of the operation while also identifying interdependency between tactical milestones; organizes a review of S-APT-related information security literature using said model; and ultimately explains ways in which S-APT operators and backing entities can be targeted for counterattack via the situation awareness dimension.

While the operational-level disinformation-based countermeasures referred to in this paper (again, see Fraunholz et al. 2018 for a thorough discussion of deception technologies intended to disorient or otherwise mislead attackers) have been employed successfully in various contexts, to our knowledge this paper is the first to examine the possible use of S-APTs as vectors for strategic disinformation as well as the potential role of information security professionals in such activities. Because S-APTs have implications that extend well beyond the immediate consequences of information security incidents, future research should be multidisciplinary in nature, examining more closely the roles APTs play within the broader strategies of nations, businesses, and criminal syndicates. The motives of these different backing entities can differ greatly.

While the usefulness of information to attackers cannot always be fully understood, because many types of information are potentially valuable for many different reasons, some types of information are valuable for predictable reasons, such as toward gaining a competitive edge in a certain industry (Ahmad et al., 2014). It would be interesting work to develop general guidelines around what kinds of strategic deception are likely to work and which are unlikely to work in the context of S-APT operations, or how different types of disinformation can be expected to impact different types of planning for different types of backing entities. While the prospect of shaping the long-term strategic agendas of these backing entities to suit one's own may not be realistic, the prospect of leveraging disinformation to gain advantage at points in time is realistic, as has been shown across various contexts for as long as there have been sentient beings capable of deception. Perhaps most importantly, the cost-benefit analysis for S-APT operations will shift dramatically if disinformation becomes a standard information security measure. Verifying the authenticity of data—whether it be the contents of obtained documents or the data accessed to induce, confirm, or measure the effectiveness of sabotage—will become costlier in terms of the time and resources involved.

This study has several limitations. First, primarily peer-reviewed sources were treated in our review of the literature. The clear majority of material published on APTs is not peer-reviewed, and it is possible that a more thorough analysis of these other bodies of literature could yield new and equally relevant insights into universal characteristics of the threat (especially where correlations between certain tools tactics, or procedures and certain types of organization may exist). Second, some limitations affect the potential applicability of the APTOL model presented in this paper. It may be possible that the role of human situation awareness is quite different for some types of S-APTs due to roles of automation or even artificial intelligence; or that such factors might sometimes drastically alter the shape of an S-APT operation (e.g. reconnaissance may be unnecessary in some situations due to superior inferential capabilities). Third, the disinformation model in this paper is based on only one theoretical model of situation awareness (with which some of the authors have worked in the past). It is possible that another theoretical model of situation awareness could be more valuable toward understanding or devising disinformation attacks on awareness and decision making in the contexts discussed within this paper.

## 10.0 Conclusion

This paper offers an 'organized conflict' perspective on APTs which supports a more holistic understanding of strategically motivated APTs as operations that are carried out to fulfill strategic objectives beyond the apparent purpose of the operation. The socio-organizational approach adopted for this research makes useful contributions to both theory and practice. The paper contributes to theory by illustrating a) how principles from military science are applicable to APTs and b) how situation awareness theory is useful toward understanding requirements for deception. The paper contributes to practice by explaining how disinformation can be used to deceive or counterattack the backing entities that sponsor and profit from strategically motivated APT operations.

## Acknowledgements

This work is supported by the Australian Research Council through the Discovery Projects scheme (DP160102277) – Enhancing Information Security Management through Organisational Learning.